\documentclass[twocolumn,epjc3]{svjour3}          % twocolumn

\RequirePackage[T1]{fontenc}

\smartqed  % flush right qed marks, e.g. at end of proof

\RequirePackage{graphicx}
\RequirePackage{mathptmx}      % use Times fonts if available on your TeX system
\RequirePackage{flushend}
\RequirePackage[numbers,sort&compress]{natbib}
\RequirePackage[colorlinks,citecolor=blue,urlcolor=blue,linkcolor=blue]{hyperref}

\journalname{Eur. Phys. J. C}

\begin{document}

\title{Gravitational Rutherford scattering and Keplerian orbits 
for electrically charged bodies in heterotic string theory}

\author{J. R. Villanueva\thanksref{e1,addr1,addr2}
        \and
        Marco Olivares\thanksref{e2,addr3} %etc.
}

%\thankstext[$\star$]{t1}{Thanks to the title}
\thankstext{e1}{e-mail: jose.villanuevalob@uv.cl}
\thankstext{e2}{e-mail: marco.olivaresr@mail.udp.cl}

\institute{Instituto de F\'{\i}sica y Astronom\'ia, 
Universidad de Valpara\'iso, Gran Breta\~na 1111,
Valpara\'iso, Chile\label{addr1}
          \and
          Centro de Astrof\'isica de Valpara\'iso, Gran Breta\~na 1111, Playa Ancha,
Valpara\'iso, Chile\label{addr2}
          \and
          Facultad de Ingenier\'ia, Universidad
Diego Portales, Avenida Ej\'ercito Libertador 441, Casilla 298--V,
Santiago, Chile\label{addr3}
}

\date{Received: XX XXXX 201X / Accepted: XX XXXX 201X}

\maketitle

%%%%%%%%%%%%%%%%%%%%%%%%%
%%%%%%%%%%%%%%%%%%%%
%%%%%%%%%%%%%

\abstract{
Properties of the  motion of electrically 
charged particles in the background of the Gibbons--Maeda--
Garfinkle--Horowitz--Strominger (GMGHS) black hole is presented 
in this paper. Radial and angular motion 
are studied analytically for different values
of the fundamental parameter. Therefore, \emph{gravitational Rutherford
scattering} and {\it Keplerian orbits} are analysed in detail. 
Finally, this paper 
complements previous work by 
Fernando for null geodesics
(Phys. Rev. D \textbf{85}: 024033, 2012), 
Olivares \& Villanueva
(Eur. Phys. J. C \textbf{73}: 2659, 2013) and Blaga
 (Automat. Comp. Appl. Math. {\bf 22}, 41 (2013); 
 Serb.\ Astron.\ J.\  {\bf 190}, 41 (2015)) for time--like geodesics.
}

%\keywords{Black Holes;  Heterotic string theory.}

\tableofcontents

\section{Introduction}
\label{intro}
The study of the motion of test particles around 
compact objects is one interesting way
to probe some phenomena related to 
classic tests of the general relativity. 
In the context of Einstein gravity, orbital precession experienced 
by the solar planets, particularly Mercury, and
the deflection of light was studied earlier by Einstein himself
\citep{Einstein15,Einstein16}
and other renowned scientists \citep{eddington20,clemence}.
Nearly a century after of them, a lot of research has been 
conducted on these tests together with other studies
(time delay, strong gravity, gravitational waves, etc.). 
Fundamentals and current advances can be found, 
for example, in \citep{shapiro90,clifford}.

In other spacetimes containing black hole solutions, 
the motion of particles  has also received 
a great deal of attention from the physics community.
For instance, a review  of neutral massive and massless particles moving
in the background of the Schwarzschild (S), Reissner-Nordstr\"{o}m (RN)
and Kerr (K) black holes can be found in \cite{chandra}. 
Furthermore, the inclusion of a cosmological constant $\Lambda$ leads to
the Kottler solution  \cite{kotler}, 
which spacetime is known as the Schwarzschild de Sitter (SdS) 
if $\Lambda> 0$, or the Schwarzschild anti-de Sitter  (SAdS)
if  $\Lambda < 0$. 
So, trajectories for neutral particles in a purely SdS
spacetime can be found in \cite{Jaklitsch,calvani,Hledik2,podolsky},
while different aspects of the motion of neutral particles
in the background of the purely SAdS spacetime has been
presented in \cite{K-W,Kraniotis,COV,Hackmann:2008zz,Hackmann:2008zza,cosv}. 
Uncharged particles in RN black hole 
with $\Lambda \neq 0$ was
studied by Stuchl\'ik \& Hled\'ik \cite{Hledik}, 
whereas circular orbits
was presented by Pugliese et al. \citep{pugliese}. Furthermore, 
Hackmann et al. \cite{hackmann} presents  analytical
solutions of the geodesic equation of massive test particles in
higher dimensional Schwarzschild, Schwarzschild anti--de Sitter, 
Reissner Nordstr\"{o}m, and Reissner Nordstr\"{o}m anti--de Sitter (RNAdS)
spacetimes and they obtained  complete solutions  and a classification
of the possible orbits in these geometries in term of
Weierstra{\ss}  functions.
Also, bounded  time-like geodesics in Kerr (K) spacetime are found in
\citep{fujita}, whereas equatorial circular motion in Kerr--de Sitter (KdS) and Kerr-Newman 
(KN) spacetimes was performed
by Stuchl\'ik \& Slany  \cite{Slany} and by Pugliese et al. \citep{pugliese13}, respectively.

There is just as much literature dealing with
alternatives theories of gravitation. 
In fact, here we mention only a few studies dealing with the motion of neutral particles.
For example, in conformal Weyl gravity,
there are many articles studying the motion of particles in addition to some observational tests
\citep{PI04,PII04,sultana10,sultana12,VOweyl}, 
 similar to those for asymptotically
Lifshitz spacetimes \citep{chen,germancito,cov21,vv,felipito}. The complete
causal structure of the Bardeen spacetime was presented in \citep{zhou}, 
the Rindler modified Schwarzschild geodesics in \citep{halisoy},
while the Shwarzschild version in gravity's Rainbow was performed by Leiva et al.
\citep{leiva}. Not least are the contributions from string theories.
Here we can mention the works of Maki \& Shiraishi \citep{maki94},
Hackmann et al. \citep{evita}, Hartmann \& Sirimachan \citep{betti} and
Bhadra \citep{bhadra}, among other authors.

The motion of electrically charged particles
represents another current line of investigation, which posits
interesting features due to the extra interaction between 
the particle and the electromagnetic field of the background.
In this sense, trajectories in a dipole magnetic field and 
in a toroidal magnetic field on the 
Schwarzschild background were performed by
Prasanna \& Varma and Prasanna \& Sengupta \citep{prasanna,prasanna2},
respectively, whereas
Dadhich et al. \citep{Dadhich} made calculations 
for trajectories on the same background when  the black hole is
immersed in an axially symmetric magnetic field (Ernst spacetime).
The general features of radial motion, motion along
the axis of symmetry and motion on the equatorial plane
in the field of rotating charged black holes
was performed in two parts by Balek, Bi\v{c}\'ak \& Stuchl\'ik
\citep{balek,balek2}.
Relativistic radial motion of electrically  charged 
particles in the field of a charged spherically symmetric
distribution of mass can be found in \citep{gladush}, whereas
classical electrically and magnetically charged
test particles in the same spacetime 
were made by Grunau and Kagramanova \citep{grunau}.
Trajectories on the RN black hole were performed by
Cohen \& Gautreau and Pugliese et al. when $\Lambda=0$ 
\citep{cohen,pugliese2}, and by Olivares et al.
when $\Lambda<0$ (RNAdS) \citep{monin}.
The motion on a rotating Kerr black hole inmersed in a 
magnetic field has been studied by  
Aliev \& \"{O}zdemir \citep{aliev} and by
 Takahashi \& Koyama \citep{takahashi},
the non-Kerr rotating version by
Abdujabbarov \citep{abdu13}, finally
the Kerr-Newmann background has been covered 
by Hackmann \& Xu \citep{evita13}.

The main goal of this paper is to work out the motion 
of electrically charged particles on the spacetime of 
a black hole coming from the heterotic string theory,
the so-called GMGHS black hole, the
causal structure of which has been
determined by Fernando for null
geodesic \citep{fernando}, and by 
Olivares \& Villanueva \citep{ov13}  
and Blaga \citep{blaga,Blaga:2014spa} for 
time-like geodesics. In this article
we use natural units with $c=1$ and $G=1$,
together with the value of the heterotic parameter
found earlier in \citep{ov13}, $\alpha=0.359$ [Km].
Therefore, in Section \ref{CBHHST} the 
charged black hole in heterotic string theory
is presented, in Section \ref{MCP} the fundamental equations of motion 
for electrically charged particles are obtained
using the Hamilton--Jacobi method.
In Section \ref{radtray} we perform 
a full analysis of the radial motion of 
test particles, and in Section
\ref{angmot} we solve  angular 
trajectories analytically 
and we study the gravitational 
scattering of Rutherford in detail. 
Finally, in section \ref{FRK} we 
conclude with general comments and
final remarks.

\section{Charged black holes in heterotic string theory}
\label{CBHHST}
The simplest 4-dimensional black hole solutions 
in heterotic sting theory,
which contain mass and electric charge,
are obtained from the
effective action \cite{sen92}
\begin{equation}\label{a0}
  \mathcal{I}_{hst}=\frac{1}{16\pi}\int d^4x\sqrt{-g} \left[R-2(\nabla \Phi)^2-e^{-2\Phi}F_{\mu\nu}F^{\mu\nu}\right],
\end{equation}
where $\Phi$ is the dilaton field, $R$ is the scalar curvature and
$F_{\mu\nu}=\partial_{\mu} A_{\nu}-\partial_{\nu} A_{\mu}$ is the Maxwell's field strength associated with
a $U(1)$ subgroup of $E_8\times E_8$ or Spin(32)%$/\mathbb{Z}_2$ 
\cite{hassan,sen92}.
The field equations associated with this action read
\begin{eqnarray}
 \label{a001}
  \nabla_{\mu}(e^{-2\Phi}F^{\mu\nu}) &=& 0, \\ \label{a002}
  \nabla^2\Phi+\frac{1}{2}e^{-2\Phi}F^2 &=& 0,
\end{eqnarray}
and
\begin{equation}\label{a003}
  R_{\mu\nu}=-2\nabla_{\mu}\Phi\,\nabla_{\nu}\Phi-2e^{-2\Phi}F_{\mu\lambda}F_{\nu}^{\lambda}
  +\frac{1}{2}g_{\mu\nu}e^{-2\Phi}F^2,
\end{equation}
and were solved by Gibbons \& Maeda \cite{gibbons88}, and independently by
Garfinkle, Horowitz \& Strominger \cite{garfinkle91}, and
thus this is known as the Gibbons--Maeda--Garfinkle--Horowitz--Strominger
(GMGHS) black hole, whose metric in the Einstein frame is
given by \cite{fernando,ov13}
\begin{equation}\label{tl1}
  ds^2=- \mathcal{F}(r) dt^2+\frac{dr^2}{ \mathcal{F}(r)}+\mathcal{R}^2(r)\,
  (d\theta^2+\sin^2\theta\,d\phi^2).
\end{equation}
Here the coordinates are defined in the ranges
$0<r<\infty$, $-\infty<t<\infty$, $0\leq \theta <\pi$, 
$0\leq \phi <2\pi$,
and the radial function $\mathcal{R}(r)$ is given by
\begin{equation}\label{tl2}
  \mathcal{R}(r)=
  \sqrt{r(r-\alpha)}, \qquad \alpha \equiv \frac{Q^2}{M},
\end{equation}
where $M$ is the ADM mass, $Q$ is the electric charge,
and $ \mathcal{F}(r)$ is the well--known lapse function 
of the Schwarzschild black hole,
\begin{equation}\label{tl3}
   \mathcal{F}(r)=1-\frac{2M}{r}= 1-\frac{r_+}{r}, \qquad r_+=2M.
\end{equation}
Since the coordinates ($t, \phi$) are cyclic in the metric
(\ref{tl1}), there are two conserved quantities
related to two Killing vectors fields:

\begin{center}
\begin{itemize}
\item the \textit{time-like Killing vector} $\xi_t= (1, 0, 0, 0)$ 
is related to the \textit{stationarity} of the metric:
$g_{\alpha \beta }\,\xi_{t}^{\alpha}\,u^{\beta}
=-\mathcal{F}(r)\,\dot{t}=-\sqrt{E}$ 
is a constant of motion which can be associated with 
the total energy of the test particles,
because this spacetime is asymptotically 
flat, and
\item the \textit{space-like Killing vector} $\xi_\phi= (0, 0, 0, 1)$  is related to the
axial symmetry of the metric:
$g_{\alpha \beta }\,\xi_{\phi}^{\alpha}\,u^{\beta}
=\mathcal{R}^{2}(r)\,\sin^2 \theta\,\dot{\phi}=L$
is a constant of motion corresponding 
to the angular momentum
of the particles moving 
in this geometry.
\end{itemize}
\end{center}
With all this, in the next section 
the basic equations governing 
the motion of charged particles 
in the spacetime generated by  
the GMGHS black hole are obtained 
by using the Hamilton--Jacobi formalism.

\section{Motion of charged particles}
\label{MCP}

\begin{figure}[!h]
 \begin{center}
  \includegraphics[width=80mm]{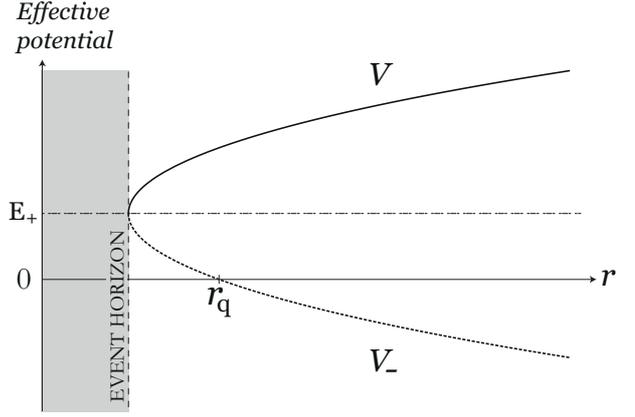}
 \end{center}
 \caption{Plot of the effective potential $V(r)$ for positive charged particles ($q>0$) 
 in radial motion together with the negative branch $V_-(r)$, which is positive in the range $r_+<r<r_q$. 
 Also, at the event horizon $r_+$, each branch converges to $E_+=q/r_+$.}
 \label{fig1a}
\end{figure}

Let us consider the motion 
of test particles which
possess mass $m$
and electric charge $\tilde{q}$. 
The Hamilton--Jacobi equation  
for the geometry described 
by the metric $g_{\mu\nu}$ is given by
\begin{equation} 
{1\over 2}g^{\mu \nu}\left(\frac{\partial S}{\partial
x^{\mu}}+\tilde{q}\, A_{\mu}\right)\left(\frac{\partial S}{\partial
x^{\nu}}+\tilde{q}\, A_{\nu}\right)+\frac{\partial S}{\partial\tau}=0,
\label{mcp1}
\end{equation}
where $S$ corresponds to the 
characteristic Hamilton function and
 $A_{\mu}$ represents the vector 
 potential components associated with
the electrodynamic properties of the black hole.
Since we are considering charged
static black holes, the only 
non-vanishing component of the vector
potential is the temporal, $A_{t}=Q/r$.
Also, the conservation 
of the angular motion implies
that the motion is developed
on an invariant plane, which 
we choose to be $\theta=\pi/2$,
so Eq. (\ref{mcp1}) reads
\begin{equation} 
\resizebox{.95\hsize}{!}{$
-\frac{1}{\mathcal{F}}\left(\frac{\partial S}{\partial t}+\frac{\tilde{q}
}{r}\sqrt{{\alpha r_+ \over 2}}\right)^{2}+\mathcal{F}\left(\frac{\partial S}{\partial
r}\right)^{2}+\frac{1}{\mathcal{R}^{2}}\left(\frac{\partial
S}{\partial \phi}\right)^{2}+2\frac{\partial S}{\partial\tau}=0.$}
\label{mpc2}
\end{equation}
Aiming to solve this last equation,  
we introduce the ansatz
$ S=-E t+S_{0}(r)+L \phi+{1\over 2}m^{2}\tau$,
together with the re-definition 
of the test charge
$q=\tilde{q}\sqrt{\frac{\alpha\, r_+}{2}}$,
and thus rewrite it as
\begin{equation} 
S_{0}(r)= \pm \int \frac{dr}{\mathcal{F}} \sqrt{(E-V_-)\,(E-V_+)},
\label{mpc3}
\end{equation} 
where the radial functions
are given by
\begin{equation} 
V_{\pm}(r)=V_q(r)\pm\sqrt{\mathcal{F} \left( m^{2}+
\frac{L^{2}}{\mathcal{R}^{2}}\right)},
\qquad V_q(r)\equiv\frac{q}{r}.
\label{mpc3.1}
\end{equation}
%%%%%%%%%%%%%%%
%%%%%%%%%%%%%%%
Notice that each branch converges to
the value $E_+=q/r_+$ at $r=r_+$, 
which can be either positive or negative, depending
on the sign of the electric charge. In  Fig. \ref{fig1a} 
we show the $q>0$ case in which 
the $V_-$ branch always is negative (except in the region $r_+<r<r_q$,
where $r_q$ is solution to the equation $V_-(r_q)=0$). 
From now on we will call the positive branch
$V_{eff}=V_{+}\equiv V$ effective potential.

Employing  the Hamilton-Jacobi method, 
it is possible to obtain three velocities 
taking into account
the motion of test particles. So, by making
${\partial S \over \partial m^2}=0$, 
${\partial S \over \partial E}=0$, and
${\partial S \over \partial L}=0$, we obtain
\begin{equation}\label{veltau}
u(r)\equiv \frac{dr}{d\tau}=\pm \sqrt{(E-V_-)\,(E-V)},
\end{equation}

\begin{equation} 
v_{t}(r)\equiv\frac{dr}{dt}=\pm  \frac{\mathcal{F}(r)\,u(r)}{E-V_{q}\left(r\right)},
\label{mpc5}
\end{equation}
and
\begin{equation}\label{velphi}
v_{\phi}(r)=\frac{dr}{d\phi}=\pm \frac{\mathcal{R}^{2}(r)\, u(r)}{L},
\end{equation}
respectively. Notice that the zeros in Eq. (\ref{veltau}),
and therefore of Eqs. (\ref{mpc5}) and (\ref{velphi}), 
correspond to the so-called turning point, $r_t$.
Furthermore, these equations lead to the quadratures
that determine the evolution of the electrically 
charged test particles, so next
sections are devoted to obtaining their analytical solutions.

\section{Radial trajectories} 
\label{radtray}
The radial motion of charged particles is characterized
by the condition $L=0$, in which case
the effective potential becomes
\begin{equation} 
V(r)=V_{q}(r)+m\sqrt{\mathcal{F}(r)}.
\label{rt1}
\end{equation}
\begin{figure}[!h]
 \begin{center}
  \includegraphics[width=80mm]{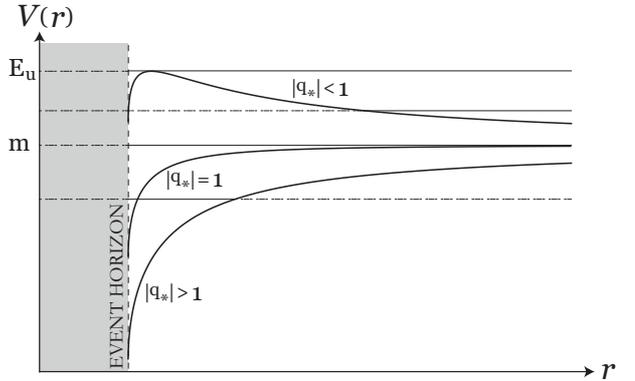}
 \end{center}
 \caption{Typical graphs of the effective potential as a function of the radial coordinate.
 Here, we show the curves in terms of the {\it electric ratio} $q_{\ast}=q_c/q$, where
 $q_c=mr_+/2$. Notice that if $|q_{\ast}|<1$ and $m<E<E_u$ a frontal scattering is 
 permitted. For $|q_{\ast}|\geq1$, the motion is essentially the same as 
 in the Schwarzschild case.}
 \label{fig1}
\end{figure}
In Fig. \ref{fig1} the effective potential
(\ref{rt1}) for three different values
of the electric charge is shown. 
A first observation of this graph is that effective 
potential presents two well defined behaviors 
in the region $r_+<r<\infty$:
\paragraph{the classic domain}
is characterized by the absence of a maximum,
so that particles with $E<m$ inexorably fall
to  the event horizon, whereas if
$E>m$, particles can escape (fall) into the spatial infinity (event horizon).
Essentially, the neutral particles exhibit this same behavior in this spacetime, 
which, as has been pointed out in \citep{ov13}, corresponds to the typical 
motion in the background of a Schwarzschild black hole. In 
Fig. \ref{fig1}, the two lower curves correspond to this domain.
\paragraph{The electric domain}
allows a maximum equal to $E_u=E_{+}\,(1+q_{\ast}^{2})$,
where  $q_{\ast}=q_c/q$ is the 
electric ratio and
$q_{c}=m\,r_+/ 2$.
Therefore, particles with $m<E<E_u$ 
feel a radial repulsion and cannot
fall into the event horizon. Furthermore, as in the previous case, 
if $E>E_u$, particles can escape (fall) into the spatial infinity (event horizon).
This domain is represented by the upper curve in Fig. \ref{fig1}.

The extreme of the effective potential is
located at
\begin{equation}\label{radmax}
\rho_{u}={r_+\over 1-q_{\ast}^{2}},
\end{equation}
so we can conclude that if
 $| q_{\ast} |<1$, then the
 position of the maximum is in the region $[r_+, \infty]$,
 while if $|q_{\ast} |>1$, then the
 location of the maximum is in the region $[-\infty, 0]$;
 finally, if  $q_{\ast}=1$, then the potential has no maximum.

Ultimately, in order to simplify the calculations,  it is
instructive rewrite square of the proper velocity
(\ref{veltau})
in the generic form:
%\begin{widetext}
\begin{eqnarray}\nonumber
u^2(r)&=&(E-V)(E-V_-)\\\nonumber
&=&\left(\frac{m^2-E^2}{r^2}\right)\left[\frac{q^2}{m^2-E^2}
+\frac{2 q (m\,q_{\ast}-E)}{m^2-E^2}r-r^2\right]\\\label{facrad}
&\equiv & \left(\frac{m^2-E^2}{r^2}\right)\,p_2(r),
\end{eqnarray}
%\end{widetext}
and thus, an exhaustive study of 
the radial motion can be carried out
taking account the values of the fundamental parameters.

\begin{figure}[!h]
 \begin{center}
  \includegraphics[width=80mm]{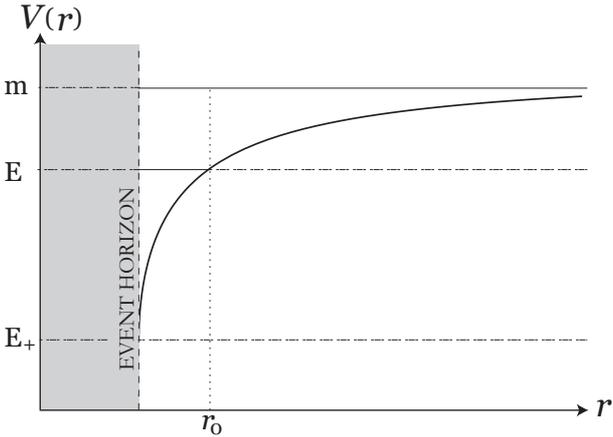}
 \end{center}
 \caption{Plot of the effective potential for radial charged particles
 in the classic domain. Bounded trajectories: particles
 with electric ratio $|q_{\ast}|> 1$ and energy $E_+<E< m$
 cannot escape to spatial infinity, 
 so the distance $r_0$ corresponds
 to a turning point. Unbounded trajectories: particles
 with electric ratio $|q_{\ast}|= 1$ and energy $m<E< E_{\infty}$
 arrive to spatial infinity with non-vanished
 kinetic energy $K\geq 0$. For simplicity, because the motion 
 is essentially the same
 for all $E \in \{m, E_{\infty}\}$, here we solve the case $E=m$ and the starting point $r_0$
 has been chosen in such way that $t(r_0)=\tau(r_0)=0$.
 }
 \label{fig2}
\end{figure}

\subsection{The classic domain}
As we have said, basically this domain is analogous
to the Schwarzschild counterpart \cite{chandra}. It is important to note 
that electric interaction is weak with respect
to gravitational effects, so charged particles 
behave like neutral particles as we have seen pointed 
out in Ref. \cite{ov13}.

\subsubsection{bounded trajectories : $|q_{\ast}|> 1$ and $E_{+}<E<m$}

\begin{figure}[!h]
 \begin{center}
  \includegraphics[width=80mm]{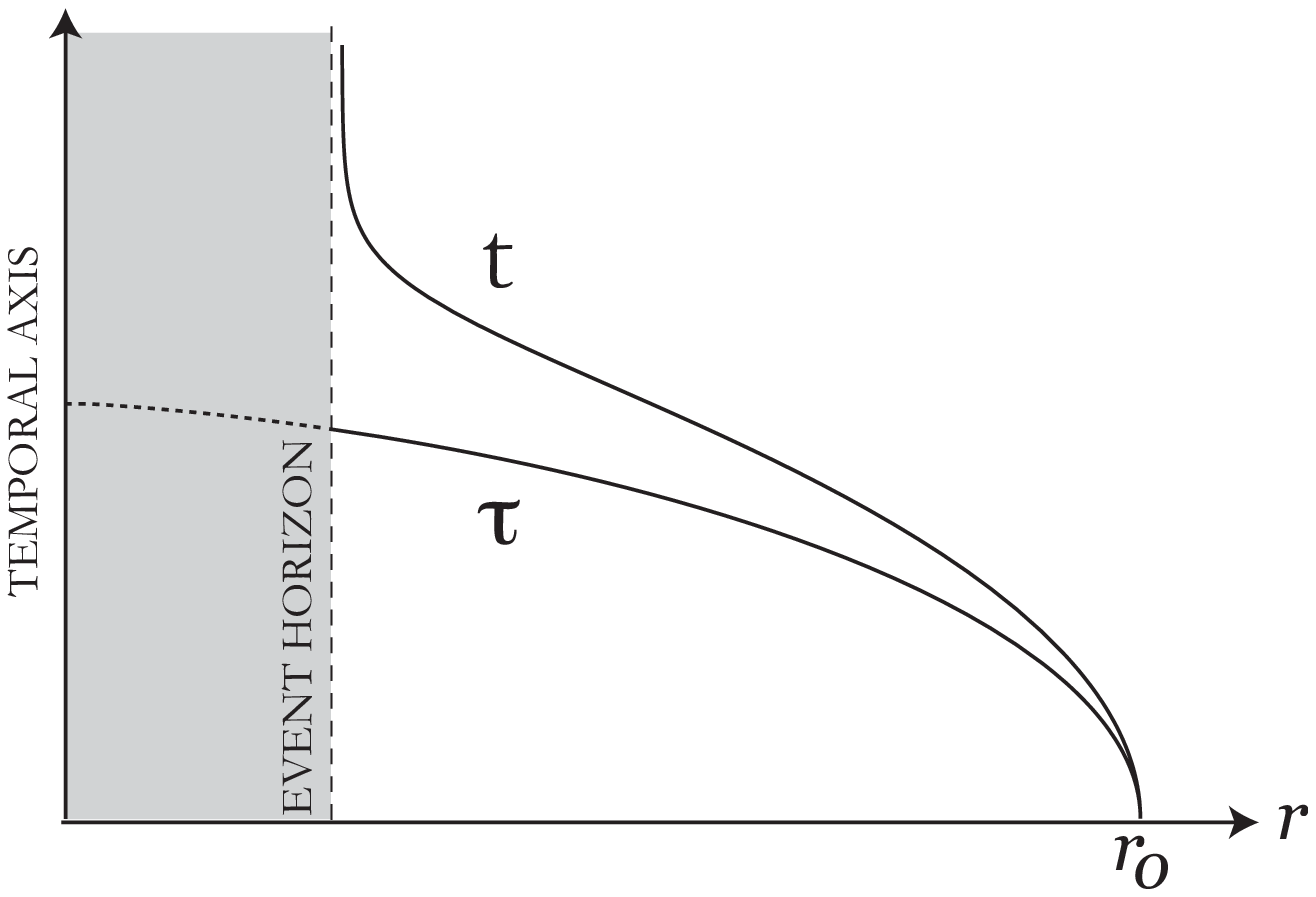}
   \includegraphics[width=80mm]{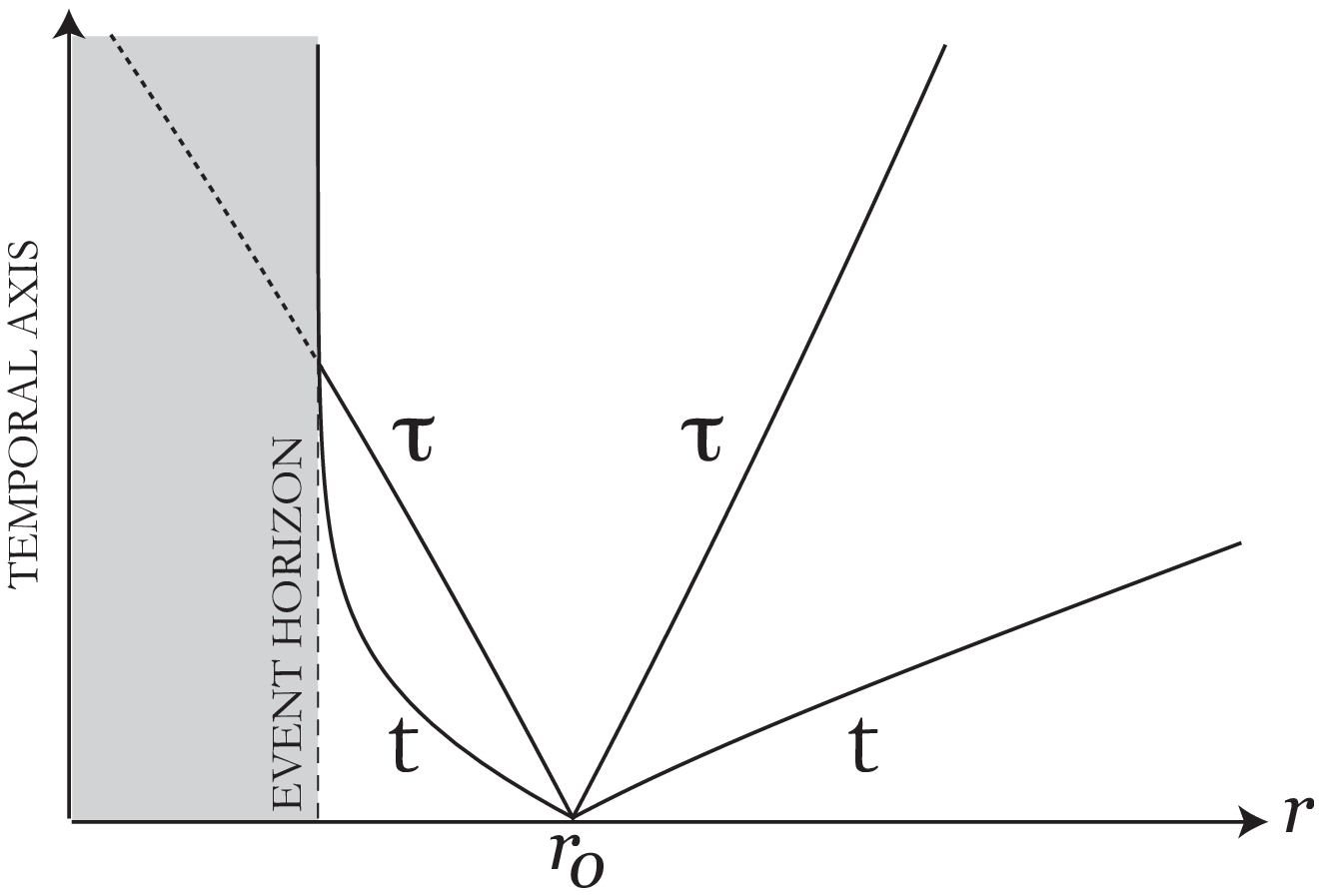}
 \end{center}
 \caption{Temporal graphic for the fall
 of radial charged particles in the classic domain.
 Basically, the  behavior is the same as
 in the standard spacetimes of general relativity \cite{chandra}.
 Top panel: Bounded trajectory with $E_+<E<m$. Bottom panel: 
 Unbounded trajectory with $E=m$. For convenience, in both graphics
 we have used the same starting point, $r_0$.}
 \label{f3}
\end{figure}

In this case, as shown in Fig. \ref{fig2}, 
we write the polynomial as $p_2(r)=(r_0-r)(r-d_0)$,
where
\begin{equation}\label{w.6}
r_0= {1\over 2}\left(b+\sqrt{b^2+4a}\right),\quad
d_0= {1 \over 2}\left(b-\sqrt{b^2+4a}\right),
\end{equation}
and, 
\begin{equation}\label{arad}
a= \frac{q^2}{m^2-E^2},\quad
b=2 q\left(\frac{m\,q_{\ast}-E}{m^2-E^2}\right).
\end{equation}
Inserting this polynomial into Eq.  (\ref{facrad}),
and then integrating Eqs. (\ref{veltau}) and  (\ref{mpc5}), we obtain
that the proper time is given by

\begin{equation}
\tau(r)={ \sqrt{ p_2(r)}+
g(0)\,\left[  \arctan\left({g(r)\over \sqrt{p_2(r)}} \right)-
\frac{\pi}{2}\right]\over  \sqrt{m^2-E^2}},
\label{taur1}
\end{equation}
while the coordinate time becomes
\begin{equation}
\label{tr1}
t(r)=\left[E+\frac{q}{g(0)}\left(\frac{E}{E_+}-1\right)\right]\,\tau(r)+\mathcal{T}(r),
\end{equation}
where
\begin{equation}
\resizebox{.95\hsize}{!}{$\mathcal{T}(r)=a \left(\frac{E}{E_+}-1\right) \left[\frac{r_+}{\sqrt{p_2(r_+)}}\log\left(\frac{ (r-r_+)\,g(r_0)}{r_+\,r
+r_0\,d_0-(r+r_+)\,g(0)}\right)-\frac{\sqrt{p_2(r)}}{g(0)}\right],$}
\end{equation}
%\begin{equation}
%\resizebox{.9\hsize}{!}{$t(r)=\left(E+\frac{r_+\,(E-E_+)}{g(0)}\right)\tau(r)+{r_+\,(E-E_+)\over \sqrt{m^2-E^2}}\left[\frac{r_+}{\sqrt{p_2(r_+)}}\log\left(\frac{ (r-r_+)\,g(r_0)}{r_+\,r
%+r_0\,d_0-(r+r_+)\,g(0)}\right)-\frac{\sqrt{p_2(r)}}{g(0)}\right],$}
%\label{tr1}
%\end{equation}
where $g(r)=\frac{1}{2}[ (r_0-r)-(r-d_0)]$, 
and we must choose 
$\tau(r_0)=t(r_0)=0$. In top panel of Fig. \ref{f3} the 
functions (\ref{taur1}) and (\ref{tr1}) are plotted together, 
demonstrating that radial charged particles 
present the same behavior as radial 
neutral particles in this spacetime
\citep{ov13}.

\subsubsection{ unbounded trajectories : $q_{\ast}=1$ and $E\geq m$ }
This case is characterized by the possibility that 
particles can arrive to spatial infinity with non-null kinetic energy, 
$K\geq 0$, where the equality is satisfied 
when $E=m$, see Fig. \ref{fig2}.
Thus, for simplicity, we solve this
last case so the proper and coordinate velocity becomes
\begin{equation}
\label{propvel1}
u(r)\equiv u_c(r)=\pm\frac{q_c}{r}, 
\end{equation}
and
\begin{equation}\label{corvel1}
v_t(r)=\pm\frac{(1-\frac{2}{m}u_c)\,u_c}{m-u_c}
\end{equation}

For simplicity, let us choose $r_0$ as the starting distance,
where  $\tau(r_0)=t(r_0)=0$; 
therefore, Eqs. (\ref{propvel1}) and (\ref{corvel1}) become
\begin{equation}
\tau(r)=\pm {r^2-r_{0}^{2}\over 2 \,q_c},
\label{mr.1}
\end{equation}
and
\begin{equation}
t(r)=m\tau(r)\pm(r-r_{0})\pm r_+\ln\left({r-r_{+}\over r_{0}-r_{+}}\right),
\label{mr.2}
\end{equation}
respectively. Therefore, as we have mentioned before, the motion
of charged particles is the same
as the motion in Einstein's spacetimes, see bottom 
panel of Fig. \ref{f3}.

\subsection{The electric domain}
\begin{figure}[!h]
 \begin{center}
  \includegraphics[width=85mm]{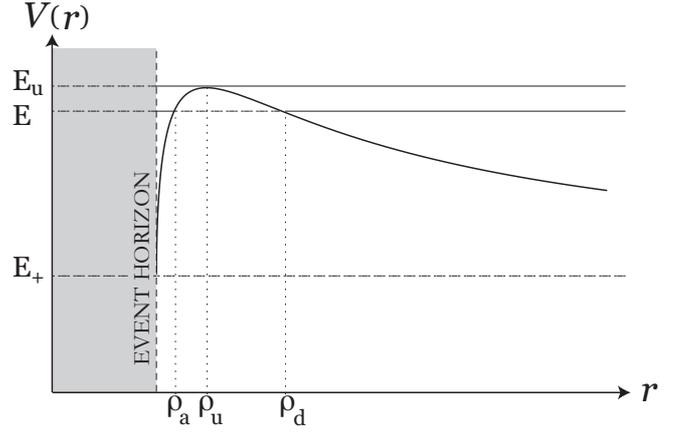}
 \end{center}
 \caption{Plot of the effective potential for 
 radial charged particles in the electric domain.
 A frontal Rutherford scattering is allowed when
 $E_+<E<E_u$ and $r>\rho_d$, while a critical radial motion
 occurs when $E=E_u$.}
 \label{f5}
\end{figure}

\subsubsection{Frontal Rutherford scattering: $|q_{\ast}|< 1$ and $m<E<E_u$}
\begin{figure}[!h]
 \begin{center}
  \includegraphics[width=85mm]{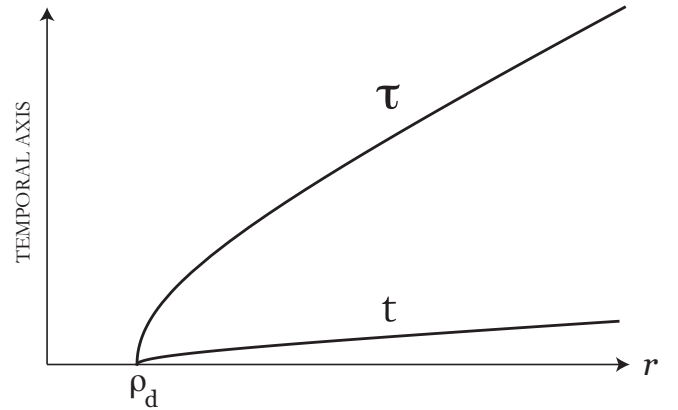}
 \end{center}
 \caption{Temporal behavior for charged particles 
 in the frontal Rutherford scattering. Here $\rho_d$ 
 represents the distance of closest approach.}
 \label{fig6}
\end{figure}

As we have said, the GMGHS spacetime
allows a Rutherford-like scattering of radial charged particles.
Thus, if the energy particle $E$ is such that 
$m<E<E_u$, then there are two
turning points, one on each side of $\rho_u$. 
Thus, the turning point 
$\rho_d>\rho_u$ corresponds
to a radial distance of closest approach
at the trajectory, whereas
$\rho_a<\rho_u$ is the farthest 
distance for this interval,
see Fig. \ref{fig1}. Explicitly, 
these distances
are given by
\begin{equation}
\rho_d= \frac{q\left(E-m\,q_{\ast}\right)}{E^{2}-m^2}
\left(1+\sqrt{1-\frac{E^{2}-m^2 }{\left(E-m\,q_{\ast}\right)^2}}\right),
\label{radruth1}
\end{equation}
and
\begin{equation}
\rho_a= \frac{q\left(E-m\,q_{\ast}\right)}{E^{2}-m^2}
\left(1-\sqrt{1-\frac{E^{2}-m^2 }{\left(E-m\,q_{\ast}\right)^2}}\right) .\label{radruth2}
\end{equation}
Therefore, Eq. (\ref{facrad}) 
can be rewritten as
\begin{equation}
\label{velpropruth}
u^2(r)=\frac{E^2-m^2}{r^2}\,h_2(r),
\end{equation}
where the polynomial is given by
$h_2(r)=|r-\rho_d|\,|r-\rho_a|$.
Assuming that $t=\tau=0$ at the turning point,
and defining the radial function $f(r)=\frac{1}{2}\left[(r-\rho_d)+(r-\rho_a) \right]$,
we found that in the proper system

\begin{equation}
\tau(r)=\frac{\sqrt{h_2(r)} +f(0) \,\log\left(\frac{\sqrt{h_2(r)} +f(r)}{\frac{1}{2}(\rho_d-\rho_a)}\right)}{ \sqrt{E^{2}-m^2}} 
\label{tauradruth1}
\end{equation}
while an external observer will measure
\begin{equation}
\label{trRu1}
t(r)=\left[E+\left(\frac{E}{E_+}-1\right)\right]\,\tau(r)+\bar{\mathcal{T}}(r),
\end{equation} 
where
\begin{equation}
\resizebox{.95\hsize}{!}{$\bar{\mathcal{T}}(r)=a \left(\frac{E}{E_+}-1\right) \left[\frac{r_+}{\sqrt{h_2(r_+)}}\log\left(\frac{ (r-r_+)\,f(\rho_d)}{r_+\,r
+\rho_d\,\rho_a-(r+r_+)\,f(0)}\right)-\frac{\sqrt{h_2(r)}}{f(0)}\right],$}
\label{trRu1}
\end{equation}
for 
$\rho_d<r<\infty$. In Fig. \ref{fig6} 
the proper and external temporal 
behavior is represented.
%, in which
%an observer at $\rho_d$ will see that test particle
%arrives to infinite in 

%\begin{equation}
%\label{tauradruth2}
%\tau(r)=\frac{\sqrt{h_2(r)}+2\,f(0)\,\textrm{arcsinh} \sqrt{\frac{r-\rho_d}{\rho_d-\rho_a}}}{\sqrt{E^{2}-m^2}}
%\end{equation}

\subsubsection{critical radial motion :  $|q_{\ast}|<1$ and $E=E_u$ }

\begin{figure}[!h]
 \begin{center}
  \includegraphics[width=80mm]{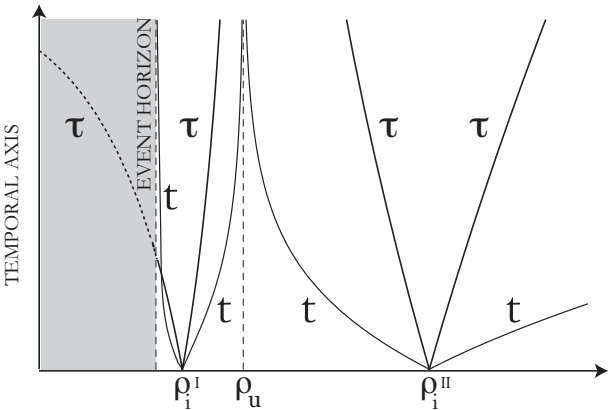}
 \end{center}
 \caption{Temporal behaviour for charged particles in critical radial motion, 
 where its energy is $E_u=V(\rho_u)$ and the starting point can be placed at left,
 $\rho_i^I$) or right of the unstable equilibrium point. Note that in both 
 frames (proper and coordinate) particles
 arrive asymptotically to $\rho_u$.}
 \label{f6}
\end{figure}

Particles with energy $E_u$ satisfying the condition
$E_u=V(\rho_u)$, where $\rho_u$ is the location
of the maximum of the effective potential, see Fig. \ref{f5}, can arrive
to $\rho_u$ either from a distance $\rho_i^{I}<\rho_u$ (region I)
or $\rho_i^{II}>\rho_u$ (region II), depending on its initial velocity.
Eventually, if the initial conditions are reversed, 
charged particles can also arrive to the spatial
infinity or the event horizon.
Under these assumptions, and then integrating 
the equations of motion, we obtain for the proper time

\begin{equation}
\tau_I (r)=\pm {1\over \sqrt{E^{2}_{u}-m^2}} \left[\rho_u \ln\,A_I(r)-(r-\rho_i^I)\right],
\label{mrc1}
\end{equation}
and
\begin{equation}
\tau_{II} (r)=\pm {1\over \sqrt{E^{2}_{u}-m^2}} \left[(r-\rho_i^{II})-
\rho_u \ln\,A_{II}(r)\right],
\label{mrc2}
\end{equation}
while the coordinate time result to be
\begin{equation}
t_I(r)=E_u\, \tau_I\pm \frac{q}{ \sqrt{E^{2}_{u}-m^2}}
\left(\frac{E_u}{E_+}-1\right)\,\ln\frac{A_I^{1+\beta}(r)}{B_I^{\beta}(r)} \label{mrc3}
\end{equation}
and
\begin{equation}
t_{II}(r)=E_u\tau_{II}(r)\mp \frac{q}{ \sqrt{E^{2}_{u}-m^2}}
\left(\frac{E_u}{E_+}-1\right)\,\ln\frac{A_{II}^{1+\beta}(r)}{B_{II}^{\beta}(r)} \label{mrc4}
\end{equation}
where we have made $\beta=r_+/(\rho_u-r_+)$, and
\begin{eqnarray}
A_I(r)&=&\frac{\rho_u-\rho_i^I }{\rho_u-r},\quad
B_I(r)=\frac{\rho_i^I-r_+}{r-r_+},\\
A_{II}(r)&=&\frac{\rho_i^{II}-\rho_u }{r-\rho_u},\quad
B_{II}(r)=\frac{\rho_i^{II}-r_+}{r-r_+},
\end{eqnarray}
In Fig. \ref{f6} the analytical solutions
(\ref{mrc1}, \ref{mrc2}, \ref{mrc3}, \ref{mrc4}) 
are depicted. Clearly, an external observer will 
see the particles going to $\rho_u$ and spatial infinity 
faster than in the proper system, while the 
motion on the horizon possesses the same nature 
as Einstein's spacetimes, i. e.,
with respect to an observer stationed
at infinite, the trajectory will take an infinite time
to reach the horizon even though by its own proper time
it will cross the horizon in a finite time \cite{chandra}.

\section{Motion with angular momentum}
\label{angmot}
\begin{figure}[!ht]
 \begin{center}
   \includegraphics[width=80mm]{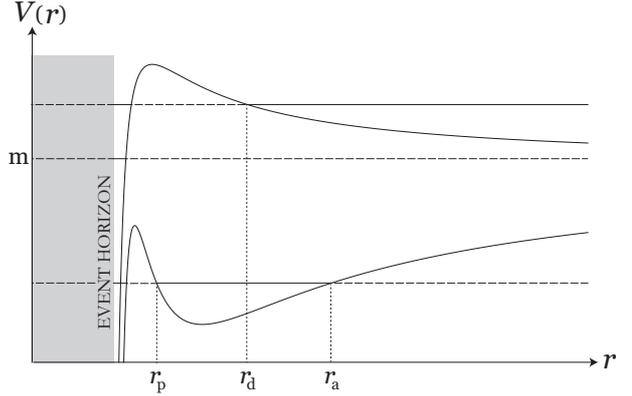}
 \end{center}
 \caption{Effective potential for charged particles with non-vanished
 angular momentum. This plot contain curves for two pairs
of value of the electric charge $q$ and angular momentum $L$
of the test particle. The upper curve corresponds to the typical 
case of dispersion in which the trajectory approaches from infinity, reaches the {\it turning point}
$r_d$, and then recedes to infinity again.}
 \label{figang}
\end{figure}

Particles with angular motion are characterized by $L > 0$.
Explicitly, the effective potential  (\ref{mpc3.1}) 
can be written as
\begin{equation} V(r)={q\over r}+  \sqrt{\left(1-{r_+\over r}\right)
\left(m^2+{L^2\over r(r-\alpha)}\right)},
\label{i.20}
\end{equation}
which is showed in Fig. \ref{figang} for two pairs
of value of the electric charge $q$ and angular momentum $L$
of the test particle.

\subsection{Gravitational Rutherford scattering}
Since the particle interacts with the background
via the presence of the term $\alpha$,
the straight path is modified
in such way that a new trajectory is
formed. As Fig. \ref{figang} illustrate,
particles with $E>m$ are deflected by
reaching the distance of closest approach 
denoted by $r_d$.
Obviously, without the $\alpha$ term
the geometry becomes the Schwarzschild one, 
and its counterpart does not exit (we put  
the interaction "off"). Moreover, a similar effect
can occur when the ratio $L/M$ increases, 
i.e., tending to the Newtonian regime
(see pp. 102 of Chandrasekhar's book \cite{chandra}).

%%%hay que decir que

In order to obtain the mentioned trajectory, 
let us rewrite Eq. (\ref{velphi}) as

\begin{equation} \phi=
{L\over \sqrt{E^2-m^2}} \int_{r_d}^{r} 
{ \,dr\over \sqrt{(r-\alpha)P(r)}}.\label{i20}\end{equation}
Here the characteristic polynomial $P(r)$ is given by
\begin{equation}\label{tl12}
P(r)=r^3- \left(r_{\sigma}+\alpha\right)r^2
-\left(r_L^2 -\alpha\,r_{\sigma} \right)r
+R_L\,r_L^2,
\end{equation}
where

\begin{equation}
\label{const1}
\resizebox{.9\hsize}{!}{$r_{\sigma}={ 2qE-m^2r_+\over E^2-m^2},
\newline 
R_L=\left({ L^2r_+-\alpha q^2\over L^2-q^2} \right),
\newline
r_L=\sqrt{{L^2-q^2\over E^2-m^2}}.$}
\end{equation}
The condition $P(r)=0$ allows three real roots, 
which can be written as

\begin{eqnarray}\label{r1rut}
r_{d}(E)&=&r_{\alpha}+R \cos \zeta,\\ \label{r2rut}
r_{A}(E)&=&r_{\alpha}-\frac{R}{2}(\cos \zeta-\sqrt{3}\sin \zeta),\\ \label{r3rut}
r_{3}(E)&=&r_{\alpha}-\frac{R}{2}(\cos \zeta+\sqrt{3}\sin \zeta),
\end{eqnarray}
where
\begin{equation}
r_{\alpha}=\frac{r_{\sigma}+\alpha}{3},\quad
R=\sqrt{\frac{\eta_2}{3}},\quad
\zeta=\frac{1}{3}\arccos \frac{\eta_3}{R^3},
\end{equation}
and
\begin{eqnarray}\label{c8}
  \eta_2  &=& 4\left[{\left(r_{\sigma}+\alpha\right)^2 \over 3}+
\left(r_L^2 -\alpha\,r_{\sigma}\right)\right], \\ \label{c9}
  \eta_3 &=& 4\left[{2 \left(r_{\sigma}+\alpha\right)^3 \over 27}+
{\left(r_{\sigma}+\alpha\right) \left(r_L^2 -\alpha\,r_{\sigma}\right)\over 3}
-R_L\,r_L^2\right].
\end{eqnarray}

\begin{figure}[!ht]
 \begin{center}
   \includegraphics[width=75mm]{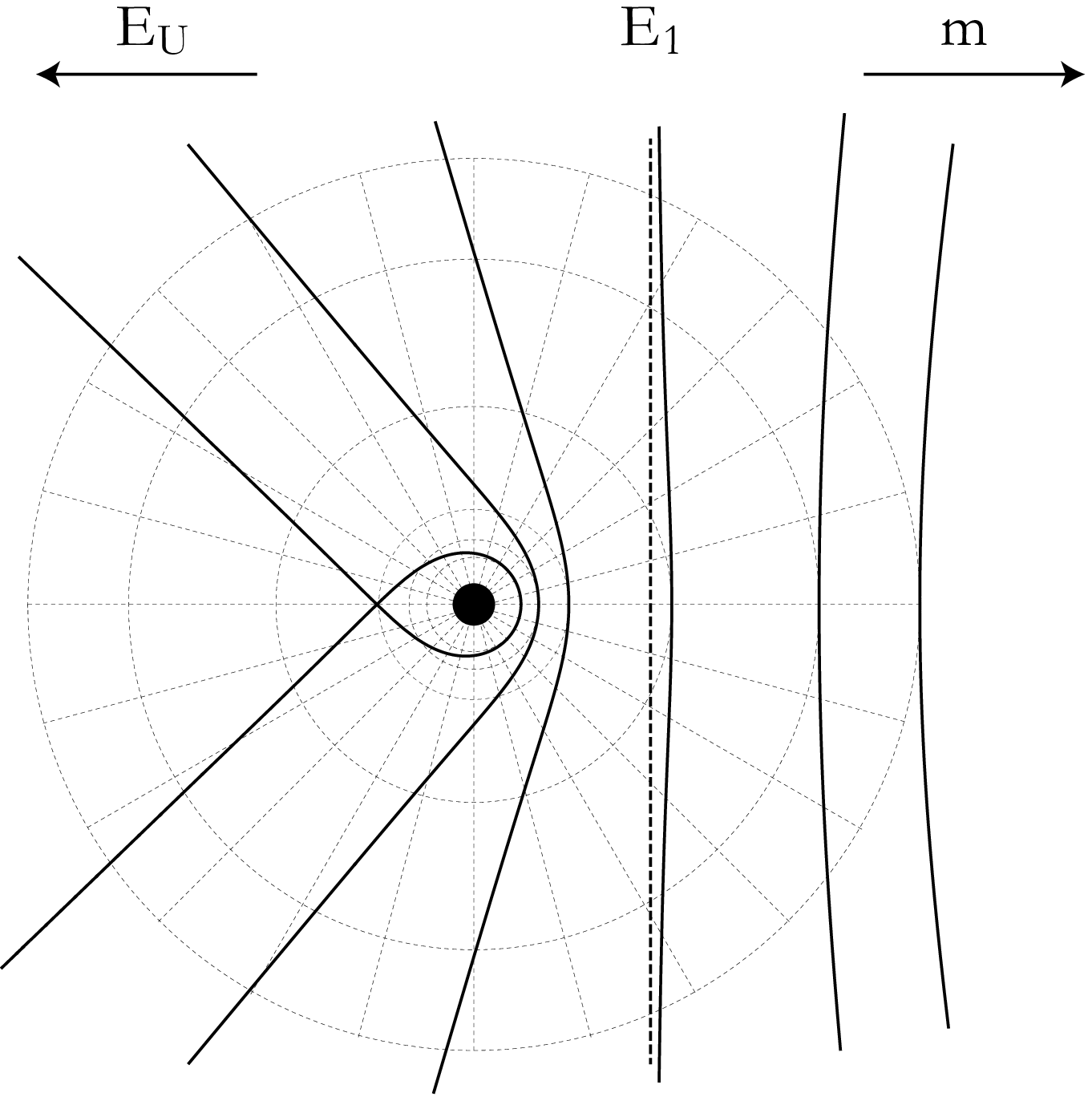}
   \includegraphics[width=75mm]{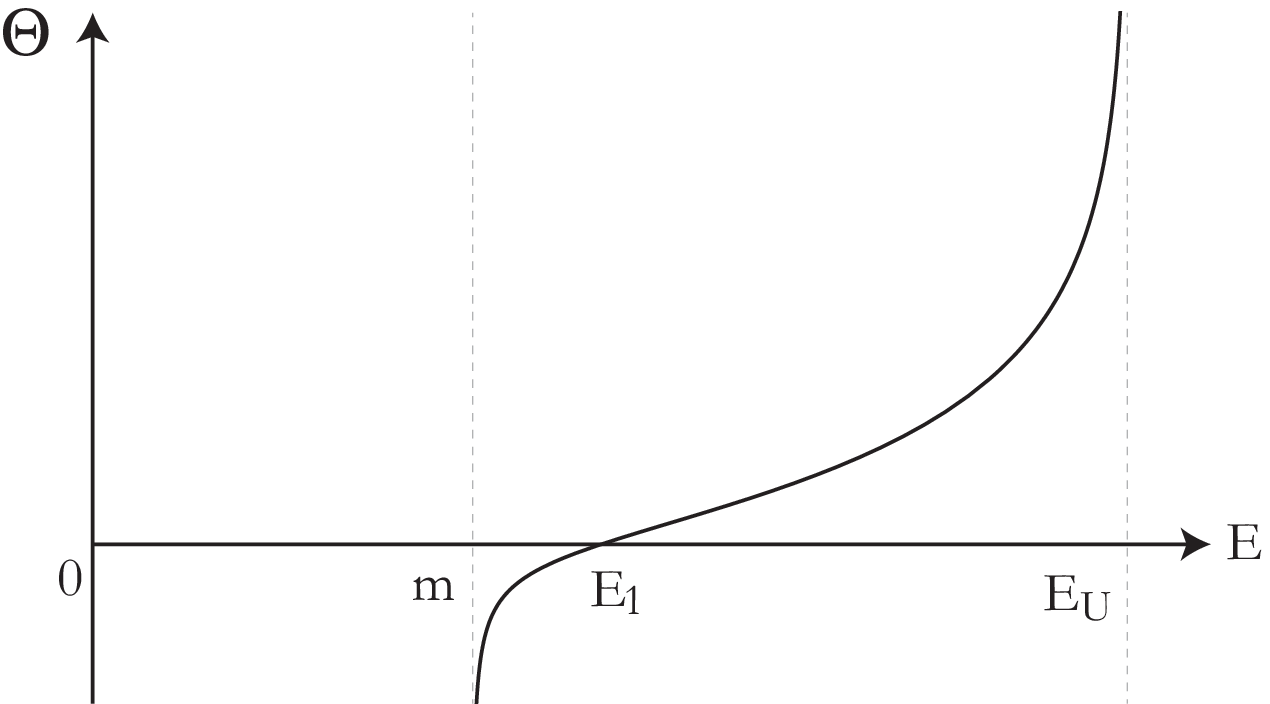}
 \end{center}
 \caption{Top panel: Gravitational Rutherford scattering. 
 This plot contain curves for various
value of the electric charge, $q$, and impact parameter, $b$,
of the test particle. Clearly, depending on the value of $b$, the scattering
can be either repulsive or attractive. Each circle
corresponds to the closest approach distance for a given value of 
the impact parameter.
Bottom panel: Angle of scattering against energy of the test particles. 
At $E=E_1$ the deflection angle is equal to zero, so initial and final direction
are the same.}
 \label{figdis}
\end{figure}

\noindent Therefore, we can identify 
the closest approach distance 
$r_{d}$, and the farthest distance $r_A$
(the third solution $r_3$ is without importance here).
Replacing $P(r)=(r-r_d)(r-r_A)(r-r_3)$
in Eq. (\ref{i20}) and performing 
an integration it is possible to find that

\begin{equation}
\label{phir1}
\kappa\,\phi=\wp^{-1}\left[\frac{1}{4}\left(\frac{1}{r-r_d}+\frac{a_1}{3}\right); g_2, g_3\right],
\end{equation}
where $\kappa =2\sqrt{E^2-m^2}/L$ 
and $\wp^{-1}(x; g_2, g_3)$ is the inverse 
$\wp$--Weierstra{\ss} elliptic function with 
the Weierstra{\ss} invariant given by

\begin{eqnarray}
\label{g2a}
g_2&=&\frac{1}{4}\left(\frac{1}{3}a_1^2-a_2\right),\\ \label{g3a}
g_3&=&\frac{1}{16}\left( \frac{1}{3} a_1\,a_2-
\frac{2}{27}a_1^3-a_3\right),
\end{eqnarray}
where $a_1=u_1+u_2+u_3$, $a_2=u_1 u_2+u_1 u_3+u_2 u_3$,
$a_3=u_1 u_2 u_3$,
with $u_1=(r_d-\alpha)^{-1}$, $u_2=(r_d-r_A)^{-1}$, and
$u_3=(r_d-r_3)^{-1}$.
Therefore, the inversion of 
Eq. (\ref{phir1}) and a brief manipulation 
leads to the following 
expression for the polar
trajectory

\begin{equation}\label{gsr}
r(\phi)= r_d+\frac{1}{4\wp(\kappa \,\phi; g_{2},g_{3})-a_1 /3},
\end{equation}

\noindent where $\wp(x; g_2, g_3)$ is the 
$\wp$--Weierstra{\ss} elliptic function.
In Fig. \ref{figdis} we plot the polar 
trajectory (\ref{gsr}), in which we
note that, depending on the set of 
parameters  ($E$, $L$, $q$, $\alpha$) across
Eqs. (\ref{const1}-\ref{c9}), the 
trajectory will be deflected
as a classic Rutherford scattering 
\cite{geiger09,geiger10,geiger13,
rutherford11,rutherford12,rutherford13,rutherford14} 
or more specifically a repulsive or 
attractive scattering between a 
massive target (composed of a positive 
charged nuclei) and light projectiles 
($\alpha$--particles or $\beta$--particles).
Moreover, because our test 
particles have been chosen as  positive, 
the attractive behavior 
is driven by the gravitational field 
over the electric repulsion.
From the equation of the orbit (\ref{phir1}) 
it is possible to obtain the angle of deflection $\Theta=
2\phi_{\infty}-\pi$
experienced by test particles, 
which turns out to be:

\begin{equation}\label{c13}
\Theta={2\over \kappa}\wp^{-1}\left({ a_1 \over 12}\right)-\pi.
\end{equation}

Therefore, particles with $E=E_1$ do not experience 
deflection in their trajectory,
where $E_1$ is the solution to the 
transcendental equation $a_1=12\,\wp(\kappa \pi/2)$.

Also, a repulsive scattering is performed if  
$m<E<E_1$, whereas 
an attractive scattering is carried out if $E_1<E<E_U$.
For the scattering problem, the 
dependence of the differential cross--section
on $\Theta\neq0$ is given by

\begin{equation}\label{seccef}
{d\,\sigma\over d\Omega}\equiv\sigma[\Theta]={b\over \sin\Theta}\left|{d\,b\over d\,\Theta}\right|,
\end{equation}
where $b$ is the impact parameter. Now, we can substitute the constants $E$
and $L$ by the value at spatial infinity as \cite{maki94}:
\begin{equation}
E={m \over \sqrt{1-v^2}},\qquad L={m\,v\,b \over \sqrt{1-v^2}}=b\,\sqrt{E^2-m^2}
\end{equation}

\noindent where $v$ is the velocity of the test
particle at spatial infinity.
Because $b=b(E)$, then $a_1=a_1(b)$ and Eq. (\ref{seccef})
is written as

\begin{equation}\label{secf}
\sigma[\Theta]={b\over \sin\Theta}\left|{d\,a_1\over d\,\Theta}\right|
\left|{d\,b\over d\,a_1}\right|.
\end{equation}

Finally, the differential cross--section for the
scattering of charged  particles by the background 
of a charged black hole in heterotic string theory is
given by

\begin{equation}\label{seft}
\sigma[\Theta]=12\,
\csc\Theta
\left|\wp^{\,\prime}\left({\pi+\Theta\over b}\right)\right|
\left|{d\,b\over d\,a_1}\right|,
\end{equation}

where $\wp^{\,\prime}(x)\equiv \wp^{\,\prime}(x, g_2, g_3)$ 
represents a derivative of the $\wp$--Weierstra{\ss} 
function with respect to $\Theta$.
Note that this last expression 
represents the exact formula
for the scattering problem. 
Nevertheless, due to the 
complexity of the relation 
between the impact parameter $b$ and 
the quantity $a_1$, the term
$\left|{d\,b\over d\,a_1}\right|$ 
is calculated numerically.

\subsection{Keplerian orbits}
As we have established, the effective 
potential (\ref{i.20}) allows motion 
between an {\it apastron} distance $r_a$ 
and {\it periastron} distance $r_p$, 
as we show in Fig.\ref{figdis}.
Therefore, considering that $E<m$, it is possible 
to rewrite Eq. (\ref{velphi}) as

\begin{equation} \phi=
{L\over \sqrt{m^2-E^2}} \int_{r_a}^{r} 
{ \,dr\over \sqrt{(r-\alpha)\bar{P}(r)}},\label{planet}\end{equation}
where the characteristic polynomial 
$\bar{P}(r)$ is now given by
\begin{equation}\label{tl12}
\bar{P}(r)=-r^3+ \left(-\bar{r}_{\sigma}+\alpha\right)r^2
-\left(\bar{r}_L^2 -\alpha\,\bar{r}_{\sigma} \right)r
+R_L\,\bar{r}_L^2,
\end{equation}

\noindent and 

\begin{equation}
\label{const1}
\resizebox{.9\hsize}{!}{$\bar{r}_{\sigma}={ 2qE-m^2r_+\over m^2-E^2},
\newline 
R_L=\left({ L^2r_+-\alpha q^2\over L^2-q^2} \right),
\newline
\bar{r}_L=\sqrt{{L^2-q^2\over m^2-E^2}}.$}
\end{equation}

The condition $\bar{P}=0$ allows us 
to get the three roots which are given by

\begin{eqnarray}
\label{rootspl1}
r_{p}(E)&=&-\bar{r}_{\alpha}+\bar{R}\sin \bar{\zeta},\\ \label{rootspl2}
r_{a}(E)&=&-\bar{r}_{\alpha}+\frac{\bar{R}}{2}(\sqrt{3}\cos \bar{\zeta}-
\sin \bar{\zeta}),\\ \label{rootspl3}
r_{f}(E)&=&-\bar{r}_{\alpha}-\frac{\bar{R}}{2}(\sqrt{3}\cos \bar{\zeta}+
\sin \bar{\zeta})
\end{eqnarray}
where
\begin{equation}
\label{angkep}
\bar{r}_{\alpha}=\frac{\bar{r}_{\sigma}-\alpha}{3},\quad
\bar{R}=\sqrt{\frac{\bar{\eta}_2}{3}},\quad\bar{\zeta}=\frac{1}{3}\arcsin
\frac{\bar{\eta}_3}{ \bar{R}^3},
\end{equation}
and
\begin{eqnarray}
\label{eta2ke}
\bar{\eta}_2&=&4\left[{\left(\bar{r}_{\sigma}-\alpha\right)^2 \over 3}-
\left(\bar{r}_L^2 -\alpha\,\bar{r}_{\sigma}\right)\right], \\ \label{eta3ke}
  \bar{\eta}_3 &=& 4\left[{2 \left(\bar{r}_{\sigma}-\alpha\right)^3 \over 27}-
{\left(\bar{r}_{\sigma}-\alpha\right) \left(\bar{r}_L^2 -\alpha\,\bar{r}_{\sigma}\right)\over 3}
-R_L\,\bar{r}_L^2\right].
\end{eqnarray}
Defining $\kappa_{kep} =2\sqrt{m^2-E^2}/L$ and then 
integrating Eq. (\ref{planet}) we get that
\begin{equation}
\label{planetang}
\kappa_{kep}\phi=\wp^{-1}\left[\frac{1}{4}\left(\frac{1}{r_a-r}-\frac{\bar{a}_1}{3}\right); \bar{g}_2, \bar{g}_3 \right],
\end{equation}
where the Weierstra{\ss} 
invariants are given by
\begin{eqnarray}
\label{gplan}
\bar{g}_2&=&\frac{1}{4}\left(\frac{1}{3}\bar{a}_1^2-\bar{a}_2\right),\\ \label{g3a}
\bar{g}_3&=&\frac{1}{16}\left( \frac{1}{3} \bar{a}_1\,\bar{a}_2-
\frac{2}{27}\bar{a}_1^3-\bar{a}_3\right),
\end{eqnarray}
where $\bar{a}_1=\bar{u}_1+\bar{u}_2+\bar{u}_3$,
$\bar{a}_2=\bar{u}_1 \bar{u}_2+\bar{u}_1 \bar{u}_3+\bar{u}_2 \bar{u}_3$,
$\bar{a}_3=\bar{u}_1 \bar{u}_2 \bar{u}_3$ and $\bar{u}_1=(r_a-\alpha)^{-1}$, 
$\bar{u}_2=(r_a-r_p)^{-1}$,  $\bar{u}_3=(r_a-r_f)^{-1}$.
Therefore, the polar trajectory can be found inverting
the Eq. (\ref{planetang}), which results to be
\begin{equation}
\label{polkep}
r(\phi)= r_a-\frac{1}{4\wp(\kappa_{kep} \,\phi; \bar{g}_{2}, \bar{g}_{3})+\bar{a}_1 /3}.
\end{equation}
\begin{figure}[!ht]
 \begin{center}
   \includegraphics[width=75mm]{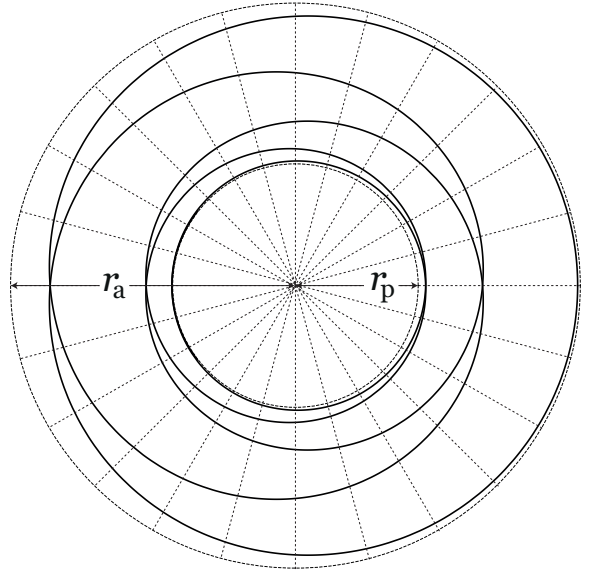}
 \end{center}
 \caption{Polar plot for Keplerian orbit performed by
 charged particles in which the precession angle is given by Eq. (\ref{angpres}).}
 \label{figplan}
\end{figure}
In Fig. \ref{figplan} we depict the Keplerian orbit
in which the precession angle, $\Xi=2\phi_p-2\pi$, is given by 
\begin{equation}
\label{angpres}
\Xi=\frac{2}{\kappa_{kep}}\,\wp^{-1}\left[\frac{1}{4}\left(\frac{1}{r_a-r_p}-\frac{\bar{a}_1}{3}\right),
\bar{g}_2, \bar{g}_3\right]-2\pi.
\end{equation}
\section{Final remarks}
\label{FRK}
In this paper, we have examined the motion
of charged particles in the background metric
and the fields of a GMGHS black hole. 
The motion equations were established and 
solved exactly following the usual 
Hamilton--Jacobi method. Thus, 
the radial motion of test bodies
was studied in terms of its 
energy $E$ and specific charge $q_{\ast}$, 
allowing two regimens: {\it the classic domain} 
is similar to the radial motion
studied in the Schwarzschild spacetimes,
thereby allowing bounded trajectories
($|q_{\ast}|>1$ and $E_+<E<m$) 
and unbounded trajectories 
($|q_{\ast}|\leq 1$ and $E\geq m$); 
and {\it the electric domain},
in which case a frontal repulsive 
Rutherford scattering is 
permitted ($|q_{\ast}|<1$
and $m<E<E_u$) together 
with a critical motion 
in which a particle {\it falls} 
asymptotically into $\rho_u$ ($|q_{\ast}|<1$
and $E=E_u$).
On the other hand, the motion 
with non--vanished 
angular momentum was 
studied in detail in 
two general schemes: 
the dispersive case $E>m$ 
and the Keplerian case $E<m$.
In the first case, we employed the classic tools
to describe the Rutherford 
scattering between two electric charges
(with the same sign of the charge), 
showing that null dispersion
and attractive scattering are 
possible because the electric dispersion
is compensated by the gravitational effects.
Finally, for the second case, 
we have calculated the apastron and 
periastron distance of the Keplerian orbit, 
which is expressed in terms of the 
elliptic $\wp$--Weierstra{\ss} function,
whose periastron advance with a 
precession angle $\Xi$ 
is given by Eq. (\ref{angpres}).
%%%%%%%%%%%%%%%%%%%%%%%%%%%%%%%%%%%%%%%%%%%%%%%%%%%%%%%%%%%%%
%%%%%%%%%%%%%%%%%%%%%%%%%%%%%5

\begin{acknowledgement}
This work was funded by the Comisi{\'o}n Nacional de 
Investigaci{\'o}n Cient{\'i}fica y Tecnol{\'o}gica through 
FONDECYT Grants No 11130695.
\end{acknowledgement}

\end{document}